\documentclass[12pt]{article}

\def\beq{\begin{equation}} \def\eeq{\end{equation}}
\def\bea{\begin{eqnarray}} \def\eea{\end{eqnarray}}

\def\beann{\begin{eqnarray*}} \def\eeann{\end{eqnarray*}}

\let\a=\alpha   
   
  \let\la=\lambda 
    
\let\om=\omega

\let\pa=\partial

\def\0{\over } \def\1{\vec }     \def\2{{1\over2}} \def\4{{1\over4}}
\def\5{\bar }  \def\6{\partial } \def\7#1{{#1}\llap{/}}

\def\<{\langle } \def\>{\rangle }

\def\sh{\mbox{\,sh}} \def\ch{\mbox{\,ch}}

\begin{document}

\newfont{\elevenmib}{cmmib10 scaled\magstep1}%
\newcommand{\tabtopsp}[1]{\vbox{\vbox to#1{}\vbox to12pt{}}}
\font\larl=cmr10 at 24pt
\newcommand{\es}{\got s}

\newcommand{\preprint}{
            \begin{flushleft}
   \elevenmib Yukawa\, Institute\, Kyoto\\
            \end{flushleft}\vspace{-1.3cm}
            \begin{flushright}\normalsize  \sf
            YITP-01-55\\
           {\tt hep-th/0108155} \\ August 2001
            \end{flushright}}
\newcommand{\Title}[1]{{\baselineskip=26pt \begin{center}
            \Large   \bf #1 \\ \ \\ \end{center}}}
\hspace*{2.13cm}%
\hspace*{0.7cm}%
\newcommand{\Author}{\begin{center}\large \bf
           V.\,I.\, Inozemtsev\footnote{
permanent address: BLTP JINR, 141980 Dubna, Moscow Region, Russia}
 and R.\, Sasaki \end{center}}
\newcommand{\Address}{\begin{center}
            Yukawa Institute for Theoretical Physics\\
     Kyoto University, Kyoto 606-8502, Japan
      \end{center}}
\newcommand{\Accepted}[1]{\begin{center}{\large \sf #1}\\
            \vspace{1mm}{\small \sf Accepted for Publication}
            \end{center}}
\baselineskip=20pt

\preprint
\thispagestyle{empty}
\bigskip
\bigskip
\bigskip

\Title{On the integrability of classical Ruijsenaars-Schneider Model
of $BC_{2}$ type}
\Author

\Address
\vspace{1cm}

\begin{abstract}
The problem of finding most general form of the classical
integrable relativistic
models of many-body interaction of the $BC_{n}$ type is considered.
In the simplest nontrivial case of $n$=2, the extra integral
of motion is presented
in explicit form  within the ansatz similar to the nonrelativistic
Calogero-Moser models. The resulting Hamiltonian has been found by
solving the set
of two functional equations.

\end{abstract}
\bigskip
\bigskip
\bigskip


\newpage
The one-parameter extensions of the Calogero-Moser integrable particle
systems first proposed
by Ruijsenaars and Schneider [1,2] have been extensively studied
during last decade [3,4,6-17].
It turned out that
both classical and quantum versions of these systems (often called
``relativistic"
but, as explained in [6], this term is misleading) were
connected with various branches
of mathematical physics. The quantum RS models with
trigonometric type of interaction
have been found to be relevant in the theory of Macdonald polynomials  and
the
integrability of their elliptic version has been investigated within
the Yang-Baxter approach [8,9].
The applications of classical models range from characteristics of
multisoliton
solutions of the sine-Gordon equation to the Seiberg-Witten theory [13].

The original $n$-particle RS model has the Hamiltonian
$$ H_{n}=\sum_{j=1}^{n}\ch p_{j}\prod_{k\neq j}^{n}F(q_{j}-q_{k}),$$
where $p,q$ are canonical variables with
Poisson brackets $\{p_{j}, p_{k}\}_{P}=\{q_{j}, q_{k}\}_{P}=0$,
$\{q_{j},p_{k}\}_{P}=\delta_{jk}$. It
is translation invariant and can be associated with $A_{n-1}$ root system.
Extensions to other root systems were proposed [7,10-17]
in ways similar to the Olshanetsky-Perelomov construction for
Calogero-Moser models [18]. One can find the first of them in the paper by
Schneider [4] who
considered the non-translation-invariant model with an external field and
the Hamiltonian
$$H_{S}=H_{n}+\sum_{j=1}^{n}W(q_{j}).$$
The Schneider model has been found to be integrable for hyperbolic choice of
the two-particle interaction,
$$F^2(q)=1+g^2\sh^2(aq), \qquad W(q)=A^2\exp(\pm 2aq),$$
which coincides upon the limit
$$p_{j}\to p_{j}/c,\quad q_{j}\to q_{j}c, \quad g\to g/c, \quad A\to
A/c,\quad c\to\infty$$
with well-known Adler system [5]. The next step has been done in the papers
by van Diejen
[7] who has studied the quantum RS systems and has found their
nine-parametric extension (with one
constraint) of the $ BC_{n}$ type in the most general elliptic case. The
results of [7] have
been proved rigorously for $n=2$ only. Komori and Hikami [10,11] also
considered the elliptic
potentials starting from the paper by Hasegawa [9]
within the framework of quantum inverse scattering method and gave a
rigorous proof of the commutativity
of van Diejen operators for arbitrary $n$ and with no constraint among
couplings. As for classical
models, the Lax operators were obtained in explicit form only for very
particular case
of $Z_{2}$ folding of the original $A_{2n}$ and $A_{2n+1}$ RS model [14-16].

The aim of this Letter is to find most general form of classical integrable
extensions
of the RS hamiltonian of the $BC$-type. The general procedure of
constructing the Lax pairs
for these systems is still lacking, so we shall consider the simplest but
nontrivial systems
of two degrees of freedom.  In this two-particle case, one needs to find
only one extra integral
of motion $J(p_{1},p_{2},q_{1},q_{2})$ so as to prove the integrability of
the system.

We shall consider the following Ansatz for the two-particle Hamiltonian and
an extra
integral of motion $J$,
\begin{equation}
H_{2}=V_{1}(q_{1}, q_{2})\ch p_{1}+V_{2}(q_{1}, q_{2})\ch
p_{2}+W(q_{1},q_{2}),  \label{eq(1)}
\end{equation}
\begin{equation}
J=\la(q_{1}, q_{2}) \ch p_{1}\ch p_{2}+\mu (q_{1},q_{2})\sh p_{1}\sh p_{2}
+\rho_{1}(q_{1},q_{2})\ch p_{1}
+\rho_{2}(q_{1},q_{2})\ch p_{2}+\tau(q_{1},q_{2}),
\label{eq(2)}
\end{equation}
where $V_{1,2}$, $W$, $\la$, $\mu$, $\rho_{1,2}$ and $\tau$ are
yet unspecified  functions of
particle coordinates. We shall suppose that $H_{2}$ and
$J$ are symmetric with respect to the
permutation of particles,
\begin{equation}
V_{2}(q_{1},q_{2})=V_{1}(q_{2}, q_{1}), \quad W(q_{1},q_{2})=W(q_{2},q_{1}),
\label{eq(3)}
\end{equation}
\begin{equation}
\la(q_{1},q_{2})=\la(q_{2},q_{1}),\quad \mu(q_{1},q_{2})
=\mu(q_{2},q_{1}),\quad
\rho_{2}(q_{1}, q_{2})=\rho_{1}(q_{2},q_{1}),\quad
\tau(q_{1},q_{2})=\tau(q_{2},q_{1}).
\label{eq(4)}
\end{equation}
The calculation of the Poisson bracket $\{H_{2},J\}_{P}$
and equating it to zero results in an overdetermined system  of 10 equations
for 8 unknown functions,
\begin{eqnarray}
V_{1}{{\pa \la}\over{\pa q_{1}}}-\la {{\pa V_{1}}\over{\pa q_{1}}}-\mu
  {{\pa V_{1}}\over{\pa q_{2}}}&=&0, \qquad
 V_{2}{{\pa \la}\over{\pa q_{2}}}-\mu {{\pa V_{2}}\over{\pa q_{1}}}-\la
  {{\pa V_{2}}\over{\pa q_{2}}}=0,
  \label{eq(5)}\\
V_{1}{{\pa \mu}\over{\pa q_{1}}}-\mu {{\pa V_{1}}\over{\pa q_{1}}}-\la
  {{\pa V_{1}}\over{\pa q_{2}}}&=&0, \qquad
  V_{2}{{\pa \mu}\over{\pa q_{2}}}-\la {{\pa V_{2}}\over{\pa q_{1}}}-\mu
  {{\pa V_{2}}\over{\pa q_{2}}}=0,  \label{eq(6)}\\
V_{1}{{\pa \rho_{1}}\over{\pa q_{1}}}-
\rho_{1}{{\pa V_{1}}\over{\pa q_{1}}}&=&0, \qquad
  V_{2}{{\pa \rho_{2}}\over{\pa q_{2}}}-
\rho_{2}{{\pa V_{2}}\over{\pa q_{2}}}=0,
\label{eq(7)}\\
V_{1}{{\pa \tau}\over{\pa q_{1}}}-V_{2}{{\pa \mu}\over{\pa q_{2}}}
   -\rho_{1}{{\pa W}\over{\pa q_{1}}}&=&0,\qquad
   V_{2}{{\pa \tau}\over{\pa q_{2}}}-V_{1}{{\pa \mu}\over{\pa q_{1}}}
   -\rho_{2}{{\pa W}\over{\pa q_{2}}}=0, \label{eq(8)}\\
V_{1}{{\pa\rho_{2}}\over{\pa q_{1}}}-\rho_{1}{{\pa V_{2}}\over{\pa q_{1}}}-
  \mu{{\pa W}\over{\pa q_{2}}}-\la{{\pa W}\over{\pa q_{1}}}&=&0, \qquad
  V_{2}{{\pa\rho_{1}}\over{\pa q_{2}}}-\rho_{2}{{\pa V_{1}}\over{\pa
q_{2}}}-
  \mu{{\pa W}\over{\pa q_{1}}}-\la{{\pa W}\over{\pa q_{2}}}=0. \label{eq(9)}
\end{eqnarray}

To solve the system (5-9), it is useful to start from four equations (5,6).
They can be
easily linearized and the general solution which takes
into account also the symmetry
(3) reads
\begin{equation}
V_{1}= F(q_{1}-q_{2})G(q_{1}+q_{2})R(q_{1}),
\qquad V_{2}=F(q_{1}-q_{2})G(q_{1}+q_{2})
R(q_{2}),\label{eq(10)}
\end{equation}
where $F,G,R$ are (still unknown) functions of one argument, and
\begin{equation}
\la=-{1\over 2}[F^2(q_{1}-q_{2})+G^2(q_{1}+q_{2})]R(q_{1})R(q_{2}), \quad
  \mu= {1\over 2}[F^2(q_{1}-q_{2})-G^2(q_{1}+q_{2})]R(q_{1})R(q_{2}).
\label{eq(11)}
\end{equation}
The solution of (\ref{eq(7)}) has the form
\begin{equation}
\rho_{1}=V_{1}d(q_{2}), \qquad \rho_{2}=V_{2}d(q_{1}).\label{eq(12)}
\end{equation}
The equations (\ref{eq(8)}) can be rewritten as
\begin{equation}
{{\pa \tau}\over{\pa q_{1}}}=
{{R(q_{2})}\over{R(q_{1})}}{{\pa \mu}\over{\pa q_{2}}}
  +d(q_{2}){{\pa W}\over{\pa q_{1}}},\qquad
  {{\pa \tau}\over{\pa q_{2}}}=
{{R(q_{1})}\over{R(q_{2})}}{{\pa \mu}\over{\pa q_{1}}}
  +d(q_{1}){{\pa W}\over{\pa q_{2}}}. \label{eq(13)}
\end{equation}
These equations with the use of (\ref{eq(11)}) can be cast into the form
\begin{equation}
{{\pa \tau}\over{\pa q_{1}}}=R(q_{2})[{1\over2}R'(q_{2})(F^2-G^2)
-R(q_{2})(FF'+GG')]
   +d(q_{2}){{\pa W}\over{\pa q_{1}}},\label{eq(14)}
\end{equation}
\begin{equation}
{{\pa \tau}\over{\pa q_{2}}}=
R(q_{1})[{1\over2}R'(q_{1})(F^2-G^2)+R(q_{1})(FF'-GG')]
   +d(q_{1}){{\pa W}\over{\pa q_{2}}},\label{eq(15)}
\end{equation}
where prime means differentiation of the function with respect to its
argument.
Now the equality of second mixed derivatives  of $\tau$ obtained from
(\ref{eq(14)},\ref{eq(15)}) yields
\begin{eqnarray}
&& {{\pa^2 W}\over {\pa q_{1}\pa q_{2}}}(d(q_{2})-d(q_{1}))
+d'(q_{2}){{\pa W}\over
{\pa q_{1}}}-d'(q_{1}){{\pa W}\over{\pa q_{2}}}\nonumber\\
&&\ \ +{{\pa}\over{\pa q_{2}}}[{1\over2}R(q_{2})R'(q_{2})
(F^2-G^2)-R^2(q_{2})(FF'+GG')]
\nonumber\\
&&\ \ -{{\pa}\over{\pa q_{1}}}[{1\over2}R(q_{1})R'(q_{1})(F^2-G^2)
+R^2(q_{1})(FF'-GG')]=0.
\label{eq(16)}
\end{eqnarray}
It is useful to introduce  instead of $F,G$ and $R$ new functions by
\begin{equation}
F=(1+f)^{1/2}, \qquad G=(1+g)^{1/2}, \qquad R=(1+r)^{1/2}. \label{eq(17)}
\end{equation}
The substitution of (\ref{eq(10)}-\ref{eq(12)}, \ref{eq(17)})
 into the last pair of equations (\ref{eq(9)}) yields after some
transformations
\begin{equation}
{{\pa W}\over{\pa q_{1}}}={1\over2}[(f'+g')(d(q_{2})-d(q_{1}))
-(1+f)(d'(q_{1})+d'(q_{2}))
  +(1+g)(d'(q_{2})-d'(q_{1}))],\label{eq(18)}
\end{equation}
\begin{equation}
{{\pa W}\over{\pa q_{2}}}={1\over2}[(f'-g')(d(q_{2})-d(q_{1}))
-(1+f)(d'(q_{1})+d'(q_{2}))
  -(1+g)(d'(q_{2})-d'(q_{1}))].\label{eq(19)}
\end{equation}
The compatibility condition for the equations (\ref{eq(18)}-\ref{eq(19)})
reads
\begin{eqnarray}
&&\hspace{-5mm}2(f''(q_{1}-q_{2})-g''(q_{1}+q_{2}))(d(q_{2})
-d(q_{1}))+ (f(q_{1}-q_{2})-g(q_{1}+q_{2}))
(d''(q_{2})-d''(q_{1}))\nonumber\\
&&\qquad\qquad -3f'(q_{1}-q_{2})(d'(q_{1})
+d'(q_{2}))+3g'(q_{1}+q_{2})(d'(q_{1})-d'(q_{2}))=0.
\label{eq(20)}
\end{eqnarray}
The equation (\ref{eq(16)}) can be transformed with
the use of (\ref{eq(17)}-\ref{eq(19)}) to the form
\begin{eqnarray}
&&\hspace{-4mm}2(f''(q_{1}-q_{2})-g''(q_{1}+q_{2}))(r(q_{2})-r(q_{1}))
  +(f(q_{1}-q_{2})-g(q_{1}+q_{2}))(r''(q_{2})-r''(q_{1}))\nonumber\\
&&\quad -3f'(q_{1}-q_{2})(r'(q_{1})+r'(q_{2}))
+3g'(q_{1}+q_{2})(r'(q_{1})-r'(q_{2}))\nonumber\\
=&&\hspace{-6mm}
(d(q_{1})-d(q_{2}))[(d''(q_{1})+d''(q_{2}))(g(q_{1}+q_{2})-f(q_{1}-q_{2}))
+3f'(q_{1}-q_{2})(d'(q_{2})-d'(q_{1}))\nonumber\\
&&\hspace{-4mm}+3g'(q_{1}+q_{2})(d'(q_{1})+d'(q_{2}))]
+2(g(q_{1}+q_{2})-f(q_{1}-q_{2}))
(d'^{2}(q_{1})-d'^{2}(q_{2})).
\label{eq(21)}
\end{eqnarray}
One is left with two functional equations
(\ref{eq(20)}, \ref{eq(21)}). The first of them allows one to find
the functions $f,g$ and $d$ while the second one can be regarded
as inhomogeneous functional equation for the function $r(q)$.
The most general solution
to (\ref{eq(20)}) with $f=g$ has been found in [19],
\begin{equation}
f(q)=g(q)=f_{0}\wp(q),\label{eq(22)}
\end{equation}
where $\wp(q)$ is the Weierstrass function
with two arbitrary periods $2\om_{1}, 2\om_{2}$,
\begin{equation}
d(q)=\sum_{i=0}^{3}\alpha_{i}\wp(q+\om_{i}), \label{eq(23)}
\end{equation}
where we use the notation $\om_{0}=0$,
$\om_{3}=-\om_{1}-\om_{2}$, and $\{\alpha_{i}\}$ are
four arbitrary real parameters.

This solution has to be used in equation (\ref{eq(21)})
which is bilinear in $d(q)$. Plugging
 (\ref{eq(22)}-\ref{eq(23)}) into right-hand side of
(\ref{eq(21)}) results, besides the terms quadratic in
$\{\a_{i}\}$, in 6 terms proportional to $\alpha_{i}\alpha_{k}$ with $i\neq
k$.
However, it is possible with the use of {\sl Mathematica}
to show that all these cross-terms
vanish and the right-hand side of (\ref{eq(21)}) acquires simple form
\[
 8\sum_{i=0}^{3}\alpha^{2}_{i}
\wp'(q_{1}+\om_{i})\wp'(q_{2}+\om_{i})[\wp(q_{2}+\om_{i})-
\wp(q_{1}+\om_{i})].
\]
It allows one to find the general solution to (\ref{eq(21)}),
\begin{equation}
r(q)=\sum_{i=0}^{3}[r_{i}
\wp(q+\om_{i})+\alpha_{i}^2\wp^{2}(q+\om_{i})],\label{eq(24)}
\end{equation}
where $\{r_{i}\}$ is yet another set of four arbitrary real parameters.
After solving equations (\ref{eq(18)}-\ref{eq(19)})
the resulting Hamiltonian reads
\begin{equation}
H_{2}=\sum_{j=1}^2
\ch p_{j}\prod_{k\neq j}\{[1+f(q_{j}-q_{k})][1+f(q_{j}+q_{k})][1+r(q_{j})]
\}^{1/2}+W(q_{1}, q_{2}),\label{eq(25)}
\end{equation}
where $f,g$ and $r$ are given by (\ref{eq(22)},\ref{eq(24)}) and
\begin{equation}
W(q_{1},q_{2})=-\sum_{i=0}^{3}[f_{0}\alpha_{i}\wp(q_{1}
+\om_{i})\wp(q_{2}+\om_{i})]-d(q_{1})-d(q_{2}).\label{eq(26)}
\end{equation}
In the papers [7,10,11] the Hamiltonian was written in terms of the products
of the Weierstrass sigma functions. Our solution (\ref{eq(24)}) for $1+r(q)$
can be also cast into this form by using the formulas
\begin{eqnarray*}
\wp(q)\wp(q+\om_{i})&=&\wp(\om_{i})[\wp(q)+\wp(q+\om_{i})]-\wp^2(\om_{i})+
\wp''(\om_{i})/2, \\[4pt]
\wp(q)-\wp(\a)&=&{{\sigma(\a-q)\sigma(\a+q)}\over
{\sigma^2 (q)\sigma^2 (\a)}}.
\end{eqnarray*}
 The second invariant $J$ is determined by the formulas
(\ref{eq(11)}-\ref{eq(12)}) and (\ref{eq(14)}-\ref{eq(15)}).
The explicit expression for the function $\tau(q_{1}, q_{2})$ can be also
obtained by the integration of (\ref{eq(14)}-\ref{eq(15)}). In the elliptic
case, it reads
\begin{eqnarray}
\tau(q_{1}, q_{2})&=&
-{1\over2}[f(q_{1}-q_{2})+f(q_{1}+q_{2})]-d(q_{1})d(q_{2})\nonumber\\
&&+f_{0}\sum_{i=0}^3[r_{i}\psi(q_{1}+\omega_{i},q_{2}+\omega_{i})
+\alpha_{i}^2\chi(q_{1}+\omega_{i}, q_{2}+\omega_{i})]\nonumber\\
&&-f_{0}\sum_{i>k=0}{{\alpha_{i}\alpha_{k}}\over
{\wp(\omega_{i}-\omega_{k})}}
\prod_{\gamma=1}^2
\wp(q_{\gamma}+\omega_{i})\wp(q_{\gamma}+\omega_{k}),\label{tauel}
\end{eqnarray}
where $f(q)$ and $d(q)$ are given by (\ref{eq(22)}-\ref{eq(23)}) and
\begin{eqnarray*}
\psi(q_{1},q_{2})&=&\wp(q_{1})\wp(q_{2})-{{\wp(q_{1})\wp'^2 (q_{2})
+\wp(q_{2})\wp'^2(q_{1})}\over {4(\wp(q_{1})-\wp(q_{2}))^2}},\\[4pt]
\chi(q_{1},q_{2})&=&-{{\wp^2 (q_{1})\wp'^2 (q_{2})+\wp^2(q_{2})\wp'^2
(q_{1})}
\over{ 4(\wp(q_{1})-\wp(q_{2}))^2}}.
\end{eqnarray*}
Taking correct hyperbolic (trigonometric) and rational limits of
(\ref{eq(22)}-\ref{eq(26)}) is a nontrivial task. To do this, we have solved
again the equations
(\ref{eq(21)}-\ref{eq(22)}) with the use of {\sl Mathematica}. The results
are given by (\ref{eq(25)}) with the following sets of functions $f,g,d,r,W$
and $\tau$:

\noindent (a) Hyperbolic case:
\begin{eqnarray}
f(q)&=&g(q)={{f_{0}}\over {\sh^2 aq}}, \label{eq(27)}\\
 d(q)&=&{{\alpha_{0}}\over {\sh^2 aq}}
+{{\alpha_{1}}\over {\ch^2 aq}}+ \alpha_{2}\ch 2aq
+\alpha_{3}\ch 4aq, \label{eq(28)}\\
r(q)&=&{{r_{0}}\over {\sh^2 aq}}
+{{r_{1}}\over{\ch^2 aq}}+r_{2}\ch 2aq +r_{3}\ch 4aq\nonumber\\
&& +{{\alpha_{0}^2}\over{\sh^4 aq}}
+{{\alpha_{1}^2}\over{\ch^4 aq}}+\alpha_{2}\alpha_{3}\ch 6aq
+{{\alpha_{3}^{2}}\over 2}\ch 8aq, \label{eq(29)}\\
W(q_{1}, q_{2})&=&-\sum_{i=1}^2 d(q_{i})-
f_{0}\left({{\alpha_{0}}\over{\sh^2 aq_{1}\sh^2 aq_{2}}}
-{{\alpha_{1}}\over{\ch^2 aq_{1}\ch^2 aq_{2}}}
+4\alpha_{3}\ch 2aq_{1}\ch 2aq_{2}\right),\nonumber\\
\label{eq(30)}
\end{eqnarray}
\vspace{-13mm}
\begin{eqnarray}
&&\tau(q_{1},q_{2})=-d(q_{1})d(q_{2})-
(\sh a(q_{1}-q_{2})\sh a (q_{1}+q_{2}))^{-2}f_{0}
\left[(\ch 2aq_{1}\ch 2aq_{2}-1)/2\right.\nonumber\\
&&\qquad\quad+r_{0}(\ch^2 aq_{1}+\ch^2 aq_{2})
+r_{1}(\sh^2  aq_{1}+\sh^2 aq_{2})
+r_{2}(\sh^2 aq_{1}\ch 2aq_{1}\ch^2 aq_{2}\nonumber\\
&&\quad+\ch^2 aq_{1}\sh^2 aq_{2}\ch 2aq_{2})+r_{3}(\sh^2 2aq_{1}\sh^2
2aq_{2}
+\sh^2 aq_{1}\ch^2 aq_{2}+\sh^2 aq_{2}\ch^2 aq_{1})\nonumber\\
&&\quad+\a_{0}^2 (\coth^2 aq_{1}+
\coth^2 aq_{2})+\a_{1}^2(\tanh^2 aq_{1}+\tanh^2 aq_{2})+\a_{2}\a_{3}
(\ch 2aq_{1}+\ch 2aq_{2})\nonumber\\
&&\qquad\qquad\times(4\ch 2aq_{1} \ch 2aq_{2}-3)
 (\ch 2aq_{1}\ch 2aq_{2}-1)/4 +\a_{3}^2(8\ch^3 2aq_{1}\ch^3
2aq_{2}\nonumber\\
&&\qquad\qquad\qquad \left.+9\ch 2aq_{1}\ch 2aq_{2}
-4\ch 2aq_{1}\ch 2aq_{2}(\ch 2aq_{1}+\ch 2aq_{2})^2
-1)/4\right]\nonumber\\
&&\qquad-f_{0}\left[16\a_{0}\a_{1}(\sh 2aq_{1}\sh 2a q_{2})^{-2}\right.
+\a_{0}\a_{2}(\sh aq_{1}\sh aq_{2})^{-2}(1+2\sh^2 aq_{1}+2\sh^2 aq_{2})
\nonumber\\
&&\qquad+\a_{1}\a_{2}(\ch aq_{1}\ch aq_{2})^{-2}(1-2\ch^2 aq_{1}-2\ch^2
aq_{2})
+\a_{0}\a_{3}(16(\sh^2 aq_{1}+\sh^2 aq_{2})\nonumber\\
&&\qquad+8(\ch^2 aq_{1}\sh^{-2} aq_{2}+\ch^2 aq_{2}\sh^{-2} aq_{1})
+(\sh aq_{1}\sh aq_{2})^{-2})
\left.+\a_{1}\a_{3}(16(\ch^2 aq_{1}\right.\nonumber\\
&&\qquad\quad\left.
+\ch^2 aq_{2})-8(\sh^2 aq_{1}\ch^{-2} aq_{2}+\sh^2 aq_{2}\ch^{-2} aq_{1})
-(\ch aq_{1}\ch aq_{2})^{-2})\right],\label{tauhyp}
\end{eqnarray}
where $\{r_{i}, \a_{i}\}$ is the set of 8 arbitrary
real parameters which does not coincide with
those introduced for general elliptic case in (\ref{eq(23)}),(\ref{eq(24)})
(the same remark holds also for subsequent
rational and asymmetric hyperbolic models).

\noindent (b) Rational case:
\begin{eqnarray}
f(q)&=&g(q)=f_{0}q^{-2}, \label{eq(31)}\\
d(q)&=&\a_{0}q^{-2}+\a_{1}q^2+\a_{2}q^4+\a_{3}q^6,\label{eq(32)}\\
r(q)&=&r_{0}q^{-2}+ r_{1}q^2 +r_{2}q^4 +r_{3}q^6 \nonumber\\
&&+\a_{0}^2 q^{-4}+(2\a_{1}\a_{3}+\a_{2}^2)q^8
+2\a_{2}\a_{3}q^{10}+\a_{3}^2 q^{12},\label{eq(33)}\\
W(q_{1}, q_{2})&=&-\sum_{i=1}^{2}d(q_{i})
-f_{0}[\a_{0}(q_{1}q_{2})^{-2}+\a_{2}(q_{1}^2 +q_{2}^2)
+\a_{3}(q_{1}^4 +q_{2}^4 +3q_{1}^2 q_{2}^2)],\label{eq(34)}\\
\tau(q_{1}, q_{2})&=&
-{1\over2}\left[f(q_{1}-q_{2})+f(q_{1}+q_{2})\right] -d(q_{1})d(q_{2})
\nonumber\\
&&\hspace{-1cm}-{{f_{0}}\over {(q_{1}^2-q_{2}^2)^2}}
\left[2r_{0}+2r_{1}q_{1}^2 q_{2}^2+r_{2}
q_{1}^2 q_{2}^2 +2r_{3}q_{1}^4 q_{2}^4 +(q_{1}^2+q_{2}^2 )(\a_{0}^2
q_{1}^{-2}q_{2}^{-2} +\a_{2}^2 q_{1}^4 q_{2}^4\right.\nonumber\\
&&\qquad \left.+\a_{3}^2 q_{1}^6 q_{2}^6
+\a_{2}\a_{3}q_{1}^4 q_{2}^4(q_{1}^2+q_{2}^2))
+\a_{1}\a_{3}q_{1}^2 q_{2}^2(q_{1}^6
+q_{2}^6+q_{1}^2 q_{2}^2(q_{1}^2+q_{2}^2))
\right]
\nonumber\\
&&-f_{0}\left[\a_{0}\a_{1}(q_{1}^{-2}+q_{2}^{-2})
+\a_{0}\a_{2}(q_{1}^2 q_{2}^{-2}
+q_{2}^2 q_{1}^{-2})\right.\nonumber\\
&&\qquad\qquad \left.+\a_{0}\a_{3}(q_{1}^4 q_{2}^{-2}+q_{2}^4 q_{1}^{-2}+
3(q_{1}^2 +q_{2}^2))
+\a_{1}\a_{2}q_{1}^2 q_{2}^2\right],\label{taurat}
\end{eqnarray}
where $f(q)$ and $d(q)$ are defined by
(\ref{eq(31)}-\ref{eq(32)}) and $\{\a_{i}, r_{i}\} $ is the
set of 8 real parameters which do not coincide with those in (\ref{tauel}).
There is yet another hyperbolic degeneration which corresponds to the
generalization of
Schneider case [4] of interaction with external field,
\begin{eqnarray}
f(q)&=&{{f_{0}}\over {\sh^2 aq}},\qquad g=0,\label{eq(35)}\\
d(q)&=&\a_{0}e^{-2aq}+\a_{1}e^{2aq}+
\a_{2}e^{-4aq}+\a_{3}e^{4aq},\label{eq(36)}\\
r(q)&=&r_{0}e^{-2aq}+r_{1}e^{2aq}+r_{2}e^{-4aq}+r_{3}e^{4aq}\nonumber\\
&&+2\a_{0}\a_{2}e^{-6aq}+ 2\a_{1}\a_{3}e^{6aq}
+\a_{2}^2 e^{-8aq}+\a_{3}^2 e^{8aq},\label{eq(37)}\\
W(q_{1}, q_{2})&=&-\sum_{i=1}^2
d(q_{i})-2f_{0}\left(\a_{2}e^{-2a(q_{1}+q_{2})}
+\a_{3}e^{2a(q_{1}+q_{2})}\right),
\label{eq(38)}\\
\tau(q_{1},q_{2})&=&-d(q_{1})d(q_{2})-f_{0}[(2\sh^2 a(q_{1}-q_{2}))^{-1}[1+
r_{0}(e^{-2aq_{1}}+e^{-2aq_{2}})/2 \nonumber\\
&&\hspace{-14mm}+r_{1}(e^{2aq_{1}}+e^{2aq_{2}})/2
+r_{2}e^{-2a(q_{1}+q_{2})}+r_{3}e^{2a(q_{1}+q_{2})}+\a_{0}\a_{2}
e^{-2a(q_{1}+q_{2})}(e^{-2aq_{1}}+e^{-2aq_{2}})\nonumber\\
&&+\a_{1}\a_{3}e^{2a(q_{1}+q_{2})}(e^{2aq_{1}}+e^{2aq_{2}})
+\a_{2}^2 e^{-4a(q_{1}+q_{2})}+\a_{3}^2 e^{4a(q_{1}+q_{2})}]\nonumber\\
&&+2\a_{0}\a_{3}(e^{2aq_{1}}+e^{2aq_{2}})
+2\a_{1}\a_{2}(e^{-2aq_{1}}+e^{-2aq_{2}})
+4\a_{2}\a_{3}\ch 2a(q_{1}-q_{2})].\label{tausch}
\end{eqnarray}
The original Schneider's result [4] for two particles corresponds to
the case of all $r_{i}$=0, $\a_{2}=\a_{3}=0$ in
(\ref{eq(35)}-\ref{eq(38)}).

Let us summarize our results.
We have demonstrated the Liouville integrability of the classical
Ruijsenaars-Schneider Hamiltonian  of the $BC_{2}$ type.
It contains nine arbitrary coupling
constants  with no constraints in the general elliptic case and its
hyperbolic and rational degenerations.
Its form (\ref{eq(25)})
is suitable for taking ``nonrelativistic" limit $p\to p/c$, $q\to cq$,
$c\to\infty$: it is necessary to rescale $f_{0}\to f_{0}/c^2$, $
r_{i}\to r_{i}/c^2, \a_{i}\to 0$ for getting most general five-parametric
two-particle
Calogero-Moser Hamiltonian given in [19].
It would be of interest to construct the Lax representation for our general
nine-parametric
case so as to prove the classical integrability for arbitrary number of
particles.
At present time it was done only for very particular cases [14-17].

{\bf Acknowledgments}. V.I.I. is supported by JSPS long term fellowship.
R.S. is partially supported by the Grant-in aid from the Ministry of
Education, Culture, Sports, Science and Technology, Japan, priority
area (\#707) ``Supersymmetry and unified theory of elementary particles".

{\bf References}
\begin{enumerate}
\item
S.N.M. Ruijsenaars and H. Schneider. Ann. Phys. (NY) {\bf 170} (1986) 370.
\item
S.N.M. Ruijsenaars. Commun. Math. Phys. {\bf 110} (1987) 191;
{\bf 115} (1988) 127.
\item
M. Bruschi and F. Calogero. Commun. Math. Phys. {\bf 109} (1987) 481.
\item
H. Schneider. Physica {\bf D26}, (1987) 203.
\item
M. Adler. Commun. Math. Phys. {\bf 55} (1977) 195.
\item
H.W. Braden and R. Sasaki. Prog. Theor. Phys. {\bf 97} (1997) 1003.
\item
J.F. van Diejen. J. Math. Phys. {\bf 35} (1994) 2983;
{\bf 36} (1995) 1299.
\item
A. Antonov, K. Hasegawa and A. Zabrodin. Nucl. Phys. {\bf B503} (1997) 747.
\item
K. Hasegawa. Commun. Math. Phys. {\bf 187} (1997) 289.
\item
Y. Komori and K. Hikami. J. Phys. {\bf A30} (1997) 434.
\item
Y. Komori and K. Hikami. J. Math. Phys. {\bf 39} (1998) 6175.
\item
K. Hasegawa, T. Ikeda and T. Kikuchi. J. Math. Phys. {\bf 40} (1999) 4549.
\item
H.W. Braden, A. Marshakov, A. Mironov, A. Morozov. Nucl. Phys. {\bf B558}
(1999) 371.
\item
B. Hou, W.-L. Yang. J. Math. Phys. {\bf 41} (2000) 357.
\item
K. Chen, B. Hou and W.-L. Yang. J. Math. Phys. {\bf 41} (2000) 8132.
\item
K. Chen and B. Hou. The $D_{n}$ Ruijsenaars-Schneider Model, hep-th/0102036.
\item
J. Avan and G. Rollet. Structures in $BC_{N}$ Ruijsenaars-Schneider
Models,\\
nlin.SI/0106015.
\item
M.A. Olshanetsky and A.M. Perelomov. Phys. Reports {\bf 71} (1981) 313.
\item
V.I. Inozemtsev. J. Phys. {\bf A17} (1984) 815.

\end{enumerate}

\end{document}